\documentstyle{mn}
\input{epsf}

\title[Evolution of spherical perturbation]{Nonlinear evolution of spherical 
perturbation in a non-flat Universe with cosmological constant}

\author[Ewa L. {\L}okas and Yehuda Hoffman]{Ewa L. {\L}okas$^1$ and Yehuda 
    Hoffman$^2$\\ $^1$Copernicus Astronomical Center, Bartycka 18,
    00--716 Warsaw, Poland\\ $^2$Racah Institute of Physics, Hebrew University, 
    Jerusalem 91904, Israel}

\begin{document}

\maketitle

\begin{abstract}
We generalize the spherical collapse model for the formation of bound objects 
to apply in a Universe with arbitrary positive cosmological constant.
We calculate the critical condition for collapse of an overdense region and
give exact values of the characteristic densities and redshifts of its
evolution. 
\end{abstract}


\section{Introduction}

The spherical collapse model was first developed by Gunn \& Gott (1972)
for a flat Universe with no cosmological constant. It assumes that the
process of formation of bound objects in the Universe can be at first
approximation described by evolution of an uniformly overdense spherical
region in otherwise smooth background (and it is therefore called the top
hat model). Despite its simplicity, the model
is still widely used to explain properties of a single bound object
via extensions such as the spherical infall model (Gunn 1977;
Hoffman \& Shaham 1985; \L okas 2000) as well as statistical
properties of different classes of objects via Press-Schechter-like
formalisms (Press \& Schechter 1974, hereafter PS; Lacey \& Cole 1993,
1994; Sheth \& Tormen 1999).

Recently, our knowledge on background cosmology has improved
dramatically mainly due to new supernovae and cosmic microwave background data.
Current observations favor an almost flat Universe with $\Omega_0=0.3$
(see e.g. Harun-or-Rashid \& Roos 2001 and references therein) and the
remaining contribution in the form of cosmological constant or some
other form of dark energy. 
A considerable effort has gone into attempts to put constraints on models
with dark energy (Wang et al. 2000; Huterer \& Turner 2001).
Another direction of investigations is into the physical basis for the
existence of such component with the oldest attempts going back to Ratra
\& Peebles (1988). One of the promising generalizations of the cosmological
constant is the quintessence model (Caldwell, Dave, \& Steinhardt 1998) 
based on so-called
``tracker fields" that display an attractor-like behavior causing the
dark energy density to follow the radiation density in the
radiation dominated era but dominate over matter density after
matter-radiation equality (Zlatev, Wang, \& Steinhardt 1999; Steinhardt,
Wang, \& Zlatev 1999). 

Given the growing popularity of models with cosmological constant we generalize 
the description of the spherical collapse to include its effect both in flat 
and non-flat cosmologies. The top hat model serves as
a basic tool in performing analytic calculations of structure formation
via gravitational instability in an expanding Universe, most notably in the
framework of the PS formalism. Our aim here is 
to extend the arsenal of analytical, or
quasi-analytical, formulae describing the redshifts and (over)densities
characterizing the collapse  processes to the case of a Universe dominated
by a cosmological constant. We derive some simple
analytical formulae and fits that will serve as useful tools in
constructing models of structure and galaxy formation.

The paper is organized as follows.
In Section~2 we briefly summarize the properties of the cosmological
model with cosmological constant including the linear growth factor of density
fluctuations.  Section~3 is devoted to the evolution of the overdense
region and gives the critical threshold for collapse.  Sections~4 and
5 discuss the characteristic densities of the forming object and
redshifts of evolution. The summary follows in Section~6.

\section{The cosmological model}

The evolution of the scale factor $a=R/R_0=1/(1+z)$ (normalized to unity
at present) in a Universe with cosmological constant is governed by the
Friedmann equation
\begin{equation}       \label{th1}
    \frac{{\rm d} a}{{\rm d} t} = \frac{H_0}{f(a)}
\end{equation}
where
\begin{equation}    \label{th2}
    f(a) = \left[ 1+ \Omega_0 \left(\frac{1}{a}
    -1\right) + \lambda_0 \left(a^2 - 1\right) \right]^{-1/2}
\end{equation}
and $H_0$ is the present value of the Hubble parameter.
The quantities with subscript $0$ here and below denote the present
values. The parameter $\Omega$ is the standard measure of the amount of
matter in units of critical density and $\lambda$
measures the density of cosmological constant in the same units
\begin{equation}    \label{q4}
    \lambda = \frac{\varrho_\Lambda}{\varrho_{\rm crit}} =\frac{\Lambda}{3 H^2}
\end{equation}
where $\Lambda={\rm const}$ is the cosmological constant in standard notation. 

The evolution of $\Omega$ and $\lambda$ with redshift $z$ is given by
\begin{equation}   \label{th3}
    \Omega(z) = \Omega_0 (1+z)^3 \left[\frac{H_0}{H(z)}\right]^2
\end{equation}
and
\begin{equation}    \label{th4}
  \lambda(z) = \lambda_0 \left[\frac{H_0}{H(z)}\right]^2 
\end{equation}
where
\begin{equation}    \label{th5}
    \left[\frac{H(z)}{H_0}\right]^2 = \Omega_0 (1+z)^3 - (\Omega_0 + \lambda_0 -1)(1+z)^2 +
    \lambda_0.
\end{equation}

The linear evolution of the matter density contrast $\delta=\delta
\varrho/\varrho$ is governed by equation
\begin{equation}    \label{th5a}
    \ddot{\delta} + 2 \frac{\dot{a}}{a} \dot{\delta} - 4 \pi G \varrho
    \delta =0
\end{equation}
where dots represent derivatives with respect to time.
In the presence of cosmological constant the growing mode can be constructed in
a simple form (Heath 1977; Carroll, Press, \& Turner 1992)
\begin{equation}    \label{th6}
    D(a) = \frac{5 \Omega_0}{2 a f(a)} \int_{0}^{a} f^3 (a) {\rm d} a
\end{equation}
where $f(a)$ was defined in equation (\ref{th2}). The
expression in (\ref{th6}) is normalized so that for $\Omega=1$ and
$\lambda=0$ we have $D(a)=a$.

For some special cases one can obtain analytical expressions for $D(a)$. In
the well-studied case of $\lambda_0=0$ and $\Omega_0 < 1$ we have
\begin{eqnarray}
    D(a) & = & \frac{a^{3/2}}{\Omega_0^{1/2}} \left[ \left(\frac{1}{a} - 1
    \right) \Omega_0 + 1 \right] ^{1/2}   \label{th8} \\
    & \times & \ _2 F_1 \left[ \frac{3}{2},
    \frac{5}{2}, \frac{7}{2}, a \left( 1-\frac{1}{\Omega_0} \right) \right]
    \nonumber
\end{eqnarray}
which is equivalent to the better known expressions given by e.g. Peebles
(1980) or Padmanabhan (1993). For $\Omega_0 + \lambda_0=1$ we get
\begin{eqnarray}
    D(a) & = & \frac{a^{3/2}}{\Omega_0^{1/2}} \left[ \left(\frac{1}{a} -
    a^2 \right) \Omega_0 + a^2 \right] ^{1/2}   \label{th9}  \\
    & \times & \ _2 F_1 \left[ \frac{5}{6},
    \frac{3}{2}, \frac{11}{6}, a^3 \left( 1-\frac{1}{\Omega_0} \right)
    \right].  \nonumber
\end{eqnarray}
For arbitrary $(\Omega_0, \lambda_0)$ pairs $D(a)$ is
easily obtained by numerical integration in equation (\ref{th6}) (see also
Hamilton 2001). 

\section{Evolution of the overdense region}

We assume that at some time $t_{\rm i}$ corresponding to redshift $z_{\rm
i}$ the region of proper radius $r_{\rm i}$ is overdense by the average
$\Delta_{\rm i} = {\rm const}$ with respect to the background, that is it
encloses a mass
\begin{equation}    \label{th10}
    M(r_{\rm i}) = \frac{4 \pi}{3} \rho_{\rm b,i} r_{\rm i}^3
    (1+\Delta_{\rm i})
\end{equation}
where $\rho_{\rm b,i}$ is the background density of matter at $t_{\rm i}$.

Evolution of this region is governed by the energy equation
\begin{equation}    \label{th11}
   \frac{1}{2} \left( \frac{{\rm d} r}{{\rm d} t} \right)^2 -
   \frac{G M}{r} - \frac{H^2 \lambda r^2}{2} = \varepsilon.
\end{equation}
Expressing $\varepsilon$ as a combination of the kinetic and potential energy per
unit mass at $t_{\rm i}$,
\begin{equation}    \label{th11a}
    \varepsilon(t_{\rm i}) = \frac{H_{\rm i}^2 r_{\rm i}^2}{2} [1 -
    \Omega_{\rm i} (1 + \Delta_{\rm i}) - \lambda_{\rm i}],
\end{equation}
using equation (\ref{th10}) and introducing a
new variable $s=r/r_{\rm i}$, equation (\ref{th11}) can be rewritten in the
form
\begin{equation}    \label{th12}
    \frac{{\rm d} s}{{\rm d} t} = \frac{H_{\rm i}}{g(a, s)}
\end{equation}
where
\begin{equation}    \label{th13} 
   g(s) = \left[ 1 + \Omega_{\rm i} (1+ \Delta_{\rm i})
    \left( \frac{1}{s} - 1 \right) + \lambda_{\rm i} (s^2 -1)
    \right]^{-1/2}.
\end{equation}
The parameters $H_{\rm i} = H(z_{\rm i})$, $\Omega_{\rm i} = \Omega(z_{\rm
i})$, $\lambda_{\rm i} = \lambda(z_{\rm i})$ are given by equations
(\ref{th3})-(\ref{th5}).

Since the energy in the overdense region is conserved we can use equations 
(\ref{th11})-(\ref{th11a}) to determine the maximum expansion (or turn-around) radius
$r_{\rm ta}$ (or equivalently, $s_{\rm ta} = r_{\rm ta}/r_{\rm i}$)
of the overdense region. It must obey the following condition
\begin{equation}    \label{th14}
    b_1 s_{\rm ta}^3 + b_2 s_{\rm ta} + b_3 = 0
\end{equation}
where
\begin{eqnarray}
    b_1 & = & \lambda_{\rm i}  \nonumber \\
    b_2 & = & 1 - \Omega_{\rm i} (1+ \Delta_{\rm
    i}) - \lambda_{\rm i}  \label{th14a}  \\
    b_3 & = & \Omega_{\rm i} (1+ \Delta_{\rm i}).  \nonumber
\end{eqnarray}

There is one interesting solution to equation (\ref{th14})
\begin{equation}     \label{th15}
     s_{\rm ta} = \frac{2}{\sqrt{3}} \left( \frac{-b_2}{b_1} \right)^{1/2}
     \cos \left(\frac{\phi - 2 \pi}{3} \right)
\end{equation}
where
\begin{equation}    \label{th16}
    \phi = \arccos \frac{x}{(x^2 + y^2)^{1/2}}
\end{equation}
with
\begin{eqnarray}
    x & = & - 9 b_1^{1/2} b_3 \label{th16a} \\
    y & = & [3 (- 4 b_2^3 - 27 b_1 b_3^2)]^{1/2}.  \label{th16b}
\end{eqnarray}
For $\lambda_0=0$ we simply get $s_{\rm ta} = -b_3/b_2$. 

The condition for the solution (\ref{th15})  to exist is
\begin{equation}    \label{th17}
    \Delta_{\rm i} \ge \Delta_{\rm i, cr} = \frac{1}{\Omega_{\rm i}}
    u(\lambda_{\rm i}) - 1
\end{equation}
where
\begin{equation}    \label{th18}
    u(\lambda_{\rm i}) = 1 + \frac{5 \lambda_{\rm i}}{4}  +
    \frac{3 \lambda_{\rm i} (8 + \lambda_{\rm i})}{4
    v(\lambda_{\rm i})} + \frac{3 v(\lambda_{\rm i})}{4}
\end{equation}
and
\begin{equation}    \label{th19}
    v = v(\lambda_{\rm i}) = \{ \lambda_{\rm i} [8 - \lambda_{\rm i}^2 + 20
    \lambda_{\rm i} + 8 (1-\lambda_{\rm i})^{3/2}] \}^{1/3}.
\end{equation}
In the limit of $\lambda_0 \rightarrow 0$ we have
$u(\lambda_{\rm i}) \rightarrow 1$. In this limit we reproduce the well
known condition for the overdense region to turn around (see e.g. Padmanabhan 1993)
\begin{equation}    \label{th20}
    \Delta_{\rm i} > \Delta_{\rm i, cr} (\lambda_0=0) =
    \frac{1}{\Omega_{\rm i}} - 1.
\end{equation}

\begin{figure}
\begin{center}
    \leavevmode
    \epsfxsize=8cm
    \epsfbox[50 50 340 310]{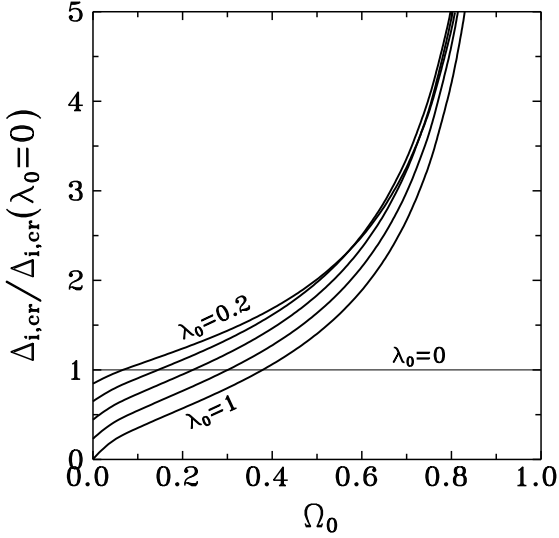}
\end{center}
    \caption{The ratio $\Delta_{\rm i, cr} /\Delta_{\rm i,
    cr} (\lambda_0=0)$ as a function of $\Omega_0$ for $z_{\rm
    i}=100$. Solid curves from top to bottom show results for
    $\lambda_0=0.2, 0.4, 0.6, 0.8$ and $1$. The thin horizontal line marks
    the limiting case of $\lambda_0=0$, where the two overdensities are
    equal.}
\label{dicr}
\end{figure}

The condition equivalent to (\ref{th17}) has been discussed previously by Weinberg (1987), 
Martel (1991) and \L okas \& Hoffman (2001).
For all positive $\lambda_0$,  we have $u(\lambda_{\rm i}) >
1$ so the condition (\ref{th17}) seems more
stringent than (\ref{th20}), i.e. in the Universe with cosmological
constant overdensities have to be larger in order to collapse than in the
Universe with the same density parameter $\Omega_{\rm i}$ but no
cosmological constant. It is more useful, however, to compare the
conditions for cosmological models with the same value of the present density
parameter, $\Omega_0$. Since the evolution of $\Omega$ depends on
$\lambda$ (see equation [\ref{th3}] and [\ref{th5}]), the relation between
conditions (\ref{th17}) and (\ref{th20}) for a given $\Omega_0$ is not
obvious. It turns out that $\Delta_{\rm i, cr}$ can in fact be higher as
well as lower than $\Delta_{\rm i, cr} (\lambda_0=0)$ depending on the
choice of $\Omega_0$ and $\lambda_0$. Figure~\ref{dicr} shows the
ratio of the two critical overdensities calculated for $z_{\rm i}=100$ as
a function of $\Omega_0$ for different values of $\lambda_0 = {\rm
const}$.

It is worth noting that in the case of $\lambda_0=0$ condition
(\ref{th20}) is equivalent to the requirement of the energy of the
overdense region to be negative. In the case of a Universe with positive
$\lambda_0$, according to equation (\ref{th11a}) the condition $E < 0$
translates into $\Delta_{\rm i} > \Omega_{\rm i}^{-1} (1-\lambda_{\rm i})
- 1$ which allows densities much lower than (\ref{th17}) since
$(1-\lambda_{\rm i}) < 1$.

\begin{figure}
\begin{center}
    \leavevmode
    \epsfxsize=8cm
    \epsfbox[50 50 340 310]{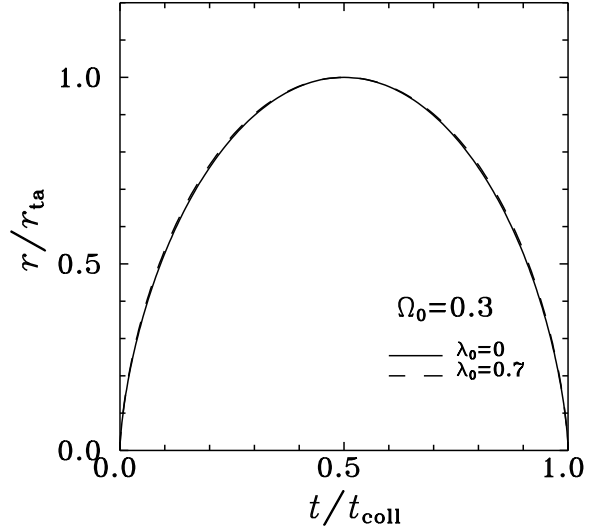}
\end{center}
    \caption{Evolution of radius of the presently collapsing overdense
    region in models with and without cosmological constant.}
\label{rodt}
\end{figure}

Integrating equation (\ref{th11}) numerically for given $\Omega_0$
and $\lambda_0$ we get the trajectory $r(t)$. Figure~\ref{rodt} compares examples of
$r(t)$ obtained for $\Omega_0=0.3$ and two values of $\lambda_0$.
For the $\Omega_0=0.3$, $\lambda_0 = 0$ case we have the well
known analytical solutions $r/r_{\rm ta} = (1-\cos \theta)/2$ and
$t/t_{\rm coll} = (\theta - \sin \theta)/(2 \pi)$ with $0 \le \theta \le 2
\pi$. The plots of $r(t)$ shown in Figure~\ref{rodt} were obtained with the
assumption of $t_{\rm coll}$ equal to the present age of Universe. In the
$\lambda_0=0$ case $r(t)$ is independent of the collapse time, while
calculations for $\lambda_0 \neq 0$ show that $r(t)$ approaches the
$\lambda_0=0$ solution for $z_{\rm coll} \rightarrow \infty$.


\section{The characteristic densities}

\subsection{The linear density contrast at collapse}

Integrating equations (\ref{th1}) and (\ref{th12}) we get
\begin{eqnarray}
    \int_0^a f(a) {\rm d} a & = & H_0 t    \label{th20a} \\
    \int_0^s g(s) {\rm d} s & = & H_{\rm i} t.   \label{th20b}
\end{eqnarray}
Eliminating $t$ we obtain equations which can be used to calculate the
scale factor at turn-around ($a_{\rm ta}$) and collapse ($a_{\rm coll}$) of
a region with particular $\Delta_{\rm i}$ at a given $z_{\rm i}$
\begin{eqnarray}
     \int_0^{a_{\rm ta}} f(a) {\rm d} a & = & \frac{H_0}{H_{\rm i}}
     \int_0^{s_{\rm ta}} g(s) {\rm d} s       \label{th21} \\
     \int_0^{a_{\rm coll}} f(a) {\rm d} a & = & 2 \frac{H_0}{H_{\rm i}}
     \int_0^{s_{\rm ta}} g(s) {\rm d} s       \label{th22}
\end{eqnarray}
where $s_{\rm ta}$ is given by equation (\ref{th15}) and we defined the
collapse time $t_{\rm coll}$ to be twice the turn-around time
\begin{equation}    \label{th21a}
    t_{\rm ta} = \frac{1}{H_0}   \int_0^{a_{\rm ta}} f(a) {\rm d} a  .
\end{equation}

Assuming that the mass inside the overdense region does not change, the
overdensity inside the sphere of size $r$ with respect to the background
density at any time is
\begin{equation}    \label{th23}
    \delta = \frac{\rho}{\rho_{\rm b}} - 1 = \frac{1}{s^3} \left(
    \frac{a}{a_{\rm i}} \right)^3 (1+ \Delta_{\rm i}) - 1
\end{equation}
where we used equation (\ref{th10}).

At early times, $t \rightarrow 0$, we can expand the expressions on the
left-hand sides of equations (\ref{th20a})-(\ref{th20b}) around $a=0$ and
$s=0$ respectively. Integrating term by term we obtain
{\samepage
\begin{eqnarray}
    H_0 t &=& \frac{2}{3 \Omega_0^{1/2}} a^{3/2} +
    \frac{\Omega_0 +\lambda_0 -1}{5 \Omega_0^{3/2}} a^{5/2} + O(a^{7/2})
    \label{th24} \\
    H_{\rm i} t &=& \frac{2}{3 [\Omega_{\rm i} (1+\Delta_{\rm
    i})]^{1/2}} s^{3/2}    \label{th25}  \\
    &+& \frac{\Omega_{\rm i} (1+\Delta_{\rm i}) +
    \lambda_{\rm i} - 1}{5 [\Omega_{\rm i} (1+\Delta_{\rm i})]^{3/2}}
    s^{5/2} + O(s^{7/2}) . \nonumber
\end{eqnarray}
}
Inverting both series we express $a$ and $s$ as power series of $t^{2/3}$
\begin{equation}    \label{th26}
    a = c_1 t^{2/3} + c_2 t^{4/3} + O(t^{8/3})
\end{equation}
where
\begin{eqnarray}
    c_1 &=& \left(\frac{3}{2} \right)^{2/3} (H_0^2 \Omega_0)^{1/3}
    \nonumber \\
    c_2 &=& \frac{3}{10} \left(\frac{3}{2} \right)^{1/3}
    \frac{H_0^{4/3} (1-\Omega_0 - \lambda_0)}{\Omega_0^{1/3}},
    \nonumber
\end{eqnarray}
and
\begin{equation}    \label{th27}
    s = d_1 t^{2/3} + d_2 t^{4/3} + O(t^{8/3})
\end{equation}
where
\begin{eqnarray}
    d_1 &=& \left(\frac{3}{2} \right)^{2/3} [H_{\rm i}^2 \Omega_{\rm i}
    (1+\Delta_{\rm i})]^{1/3} \nonumber \\
    d_2 &=& \frac{3}{10}
    \left(\frac{3}{2} \right)^{1/3} \frac{H_{\rm i}^{4/3} [1 - \Omega_{\rm
    i} (1+\Delta_{\rm i}) - \lambda_{\rm i}]}{[\Omega_{\rm i}
    (1+\Delta_{\rm i})]^{1/3}}. \nonumber
\end{eqnarray}

Inserting expressions (\ref{th26}) and (\ref{th27}) into equation
(\ref{th23}) and keeping only the lowest order term we find
\begin{eqnarray}
    \delta &=& \frac{3}{5} \left( \frac{3}{2} \right)^{2/3}
    \left[ \frac{H_0^{2/3} (1- \Omega_0 -
    \lambda_0)}{\Omega_0^{2/3}}   \right.           \label{th28}   \\
    &+& \left.  \frac{H_{\rm i}^{2/3}[\Omega_{\rm i} (1+ \Delta_{\rm i})
    + \lambda_{\rm i} -1] }{[\Omega_{\rm i} (1+\Delta_{\rm
    i})]^{2/3}}  \right] t^{2/3} + O(t^{4/3}).  \nonumber
\end{eqnarray}
From expansion (\ref{th26}) we have to the lowest order $t^{2/3} =
c_1^{-1} a$ so the dependence of $\delta$ on $a$ in the limit of $a
\rightarrow 0$ is
\begin{equation}    \label{th29}
    \delta =  h(\Omega_0, \lambda_0, \Delta_{\rm i}, z_{\rm i}) a + O(a^2),
\end{equation}
where
\begin{eqnarray}
    h(\Omega_0, \lambda_0, \Delta_{\rm i}, z_{\rm i}) &=&
    \label{th29a} \\
    && \hspace{-1in} = \frac{3}{5}
    \left[ \frac{1-\Omega_0 -\lambda_0}{\Omega_0} +
    \frac{[\Omega_{\rm i} (1+\Delta_{\rm i}) + \lambda_{\rm i}
    -1](1+z_{\rm i})}{\Omega_{\rm i} (1+\Delta_{\rm i})^{2/3}} \right] .
    \nonumber
\end{eqnarray}
Using equations (\ref{th3})-(\ref{th5}) we can rewrite $h$ in terms of the present
cosmological parameters
\begin{eqnarray}
    h(\Omega_0, \lambda_0, \Delta_{\rm i}, z_{\rm i}) &=&
    \label{th29b} \\
    && \hspace{-1in} = \frac{3}{5}
    \left[ \frac{1-\Omega_0 - \lambda_0}{\Omega_0} +
    \frac{\Omega_0 [1+\Delta_{\rm i} (1+z_{\rm i})] + \lambda_0-1}
    {\Omega_0 (1+\Delta_{\rm i})^{2/3}} \right] .
    \nonumber
\end{eqnarray}

For arbitrary $(\Omega_0, \lambda_0)$, one can show that
\begin{equation}    \label{th7}
    D(a) = a + O(a^2)
\end{equation}
is the properly normalized solution of equation (\ref{th5a}) around $a=0$.
Given this behavior of the linear growth factor $D(a)$, we finally obtain
the density contrast as predicted by linear theory
\begin{equation}    \label{th30}
    \delta_{\rm L} = h(\Omega_0, \lambda_0, \Delta_{\rm i},
    z_{\rm i}) D(a).
\end{equation}
Note that $D(a)$ is the general solution to equation (\ref{th5a}). The first order
expansion of $D(a)$ has been used only to derive $h(\Omega_0, \lambda_0,
\Delta_{\rm i}, z_{\rm i})$.

A particularly useful quantity is the linear density contrast at the
moment of collapse i.e. when $s$ reaches zero
\begin{equation}    \label{th31}
    \delta_{\rm c} = h[\Omega_0, \lambda_0, \Delta_{\rm i}(a_{\rm coll}),
    z_{\rm i}] D(a_{\rm coll}).
\end{equation}
$\Delta_{\rm i}(a_{\rm coll})$ in the above equation means that
$\Delta_{\rm i}$ corresponding to $a_{\rm coll}$ has to be determined
for a given $z_{\rm i}$ from equation (\ref{th12}). The problem can be 
reduced to solving equation (\ref{th22}) with $s_{\rm ta}$
given by equation (\ref{th15}). 

\begin{figure}
\begin{center}
    \leavevmode
    \epsfxsize=8cm
    \epsfbox[50 50 340 570]{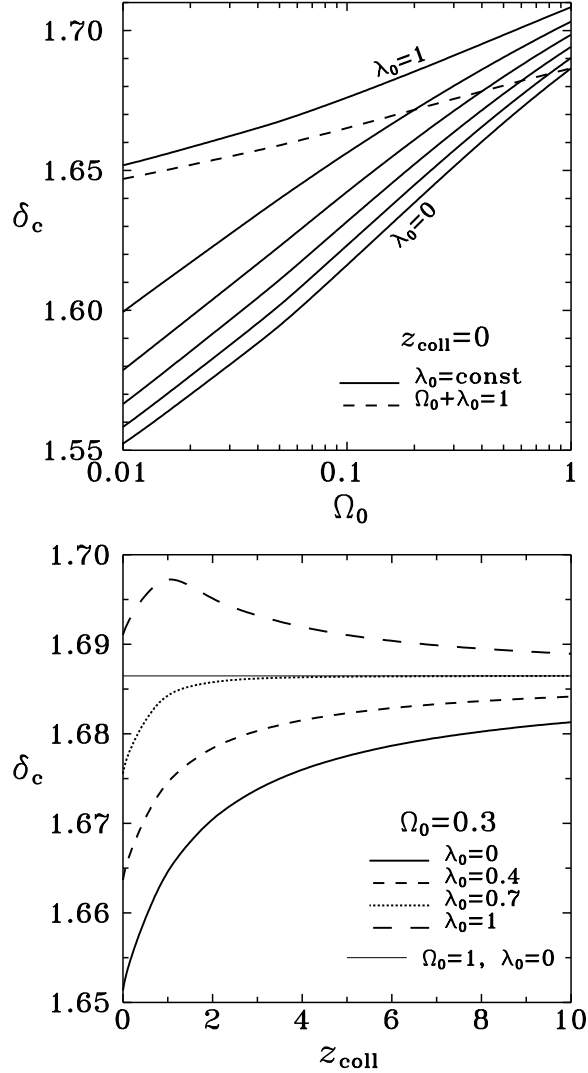}
\end{center}
    \caption{Upper panel: parameter $\delta_{\rm c}$ as a
    function of $\Omega_0$ with solid lines corresponding from bottom to
    top to $\lambda_0=0, 0.2, 0.4, 0.6, 0.8, 1$ and the dashed line
    showing results for the flat case $\Omega_0 + \lambda_0 = 1$. Lower
    panel: $\delta_{\rm c}$ as a function of $z_{\rm coll}$ for four models with
    $\Omega_0=0.3$ and $\lambda_0=0, 0.4, 0.7, 1$. The thin straight
    line marks the fiducial value $\delta_{\rm c} = 1.68647$ for $\Omega_0
    = 1$, $\lambda_0 = 0$.}
\label{dellam}
\end{figure}

Figure~\ref{dellam} shows $\delta_{\rm c}$ for different values of $\lambda_0$. 
In the upper
panel the assumption is that the collapse is taking place now, $z_{\rm
coll} = 0$. Solid lines display the dependence of $\delta_{\rm c}$ on
$\Omega_0$ with $\lambda_0 = {\rm const}$ while the dashed line has
$\lambda_0$ chosen so that $\Omega_0 + \lambda_0 = 1$. In this last
special case we reproduce the results of Eke, Cole \& Frenk (1996), while
for open models with no cosmological constant our results match those
of Lacey \& Cole (1993). 

The lower panel of Figure~\ref{dellam} shows how $\delta_{\rm c}$ changes with
the redshift of collapse, $z_{\rm coll}$, for a few models with
$\Omega_0=0.3$ and different values of $\lambda_0$. We see that the
dependence on $z_{\rm coll}$ is rather weak and at large $z_{\rm coll}$
the values converge to the well known fiducial value of $\delta_{\rm c} =
3 (12 \pi)^{2/3}/20 \approx 1.68647$ valid in the Universe with
$\Omega_0=1$ and $\lambda_0 = 0$. This value is particularly quickly
reached with increasing $z_{\rm coll}$ for the flat case, $\Omega_0=0.3$,
$\lambda_0=0.7$.

\subsection{The density of virialized halo}

Another useful quantity is the ratio of the density in the object to the
critical density at virialization
\begin{equation}    \label{th32}
    \Delta_{\rm c} = \frac{\rho_{\rm vir}}{\rho_{\rm crit}} (a_{\rm coll})
    =  \frac{\Omega (a_{\rm coll})}{s_{\rm coll}^3} \left( \frac{a_{\rm
    coll}}{a_{\rm i}} \right)^3 [1+\Delta_{\rm i} (a_{\rm coll})]
\end{equation}
where $s_{\rm coll} = r_{\rm coll}/r_{\rm i}$ and $r_{\rm coll}$ is the
effective final radius of the collapsed object. We assume that the object
virializes at $t_{\rm coll}$, the time corresponding to $s \rightarrow
0$. Application of the virial theorem in the presence of cosmological constant
leads to the following equation for the ratio of the final radius of the
object to its turn-around radius $F=r_{\rm coll}/r_{\rm ta}$ (Lahav et al.
1991)
\begin{equation}   \label{th33}
    2 \eta F^3 - (2 + \eta) F + 1 = 0
\end{equation}
where
\begin{equation}   \label{th34}
    \eta = \frac{2 \lambda_{\rm i} s_{\rm ta}^3}{\Omega_{\rm i}
    (1+\Delta_{\rm i})} .
\end{equation}
In the calculations of $\Delta_{\rm c}$ we use the exact solution to
equation (\ref{th33}) which in the case of $\lambda > 0$  can be written
down using expression (\ref{th15}) with $F$ instead of
$s_{\rm ta}$ and $b_1 = 2 \eta$, $b_2 = -(2+\eta)$ and $b_3 = 1$. However,
a good approximation is provided by $F \approx (1-\eta/2)/(2-\eta/2)$
(Lahav et al. 1991).

\begin{figure}
\begin{center}
    \leavevmode
    \epsfxsize=8cm
    \epsfbox[50 50 340 570]{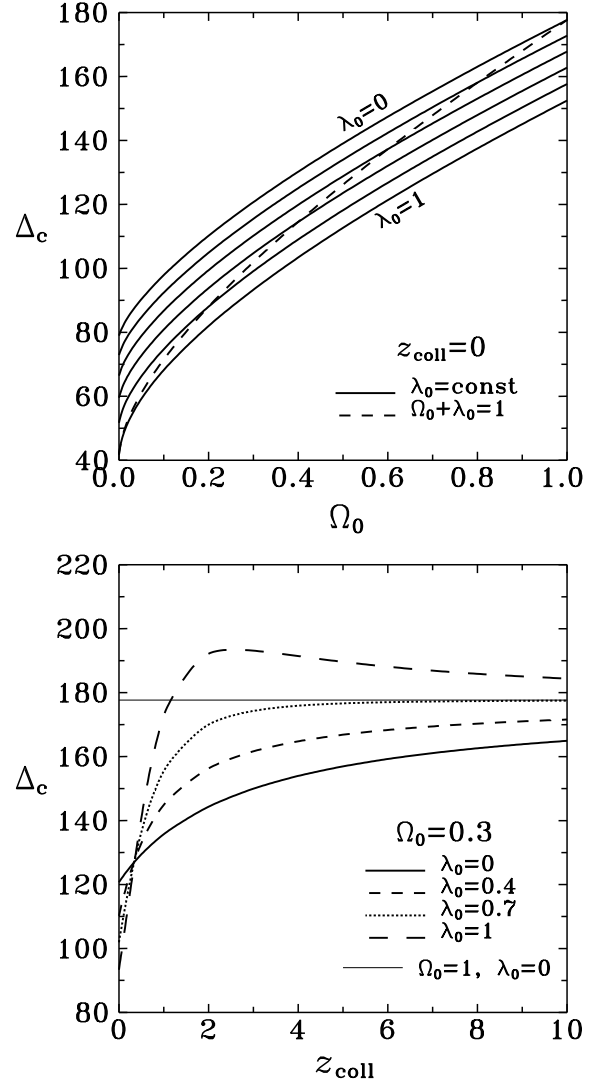}
\end{center}
    \caption{Upper panel: parameter $\Delta_{\rm c}$ as a
    function of $\Omega_0$ with solid lines corresponding from top to
    bottom to $\lambda_0=0, 0.2, 0.4, 0.6, 0.8, 1$ and the dashed line
    showing results for the flat case $\Omega_0 + \lambda_0 = 1$. Lower
    panel: $\Delta_{\rm c}$ as a function of $z_{\rm coll}$ for four models with
    $\Omega_0=0.3$ and $\lambda_0=0, 0.4, 0.7, 1$. The thin straight
    line marks the fiducial value $\Delta_{\rm c} = 177.653$ for $\Omega_0
    = 1$, $\lambda_0 = 0$.}
\label{dllam}
\end{figure}

Figure~\ref{dllam} shows $\Delta_{\rm c}$ for different $\lambda_0$. 
The upper panel gives its
values as a function of $\Omega_0$ for models with different values of
$\lambda_0$ with the assumption that the collapse occurs at $z_{\rm
coll} = 0$. The solid lines correspond to constant values of $\lambda_0$
while the dashed line has $\lambda_0$ chosen so that $\Omega_0 +
\lambda_0=1$. Again, we agree with the results for the special cases of
$\Omega_0<1$, $\lambda_0=0$ and $\Omega_0 + \lambda_0=1$ derived
previously by Lacey \& Cole (1993) and Eke et al. (1996) respectively.

We see that for a given $\Omega_0$ lower
$\lambda_0$ makes virialized objects denser with respect to critical
density. It is interesting to note, however, that this relation is
inverted for higher $z_{\rm coll}$ as proved by the lower panel
of Figure~\ref{dllam}, where we display the dependence of $\Delta_{\rm c}$ on the
redshift of collapse for a few models. Again, as in the case of
$\delta_{\rm c}$, we observe that at high $z_{\rm coll}$ values of
$\Delta_{\rm c}$ approach the well known fiducial value of $\Delta_{\rm c}
= 18 \pi^2 \approx 177.653$ valid for $\Omega_0 = 1$, $\lambda_0 = 0$ and
the convergence is fastest for the flat case.

\section{Characteristic redshifts}

It is sometimes useful to be able to estimate the redshift of a particular
stage of evolution of the perturbation given its present overdensity as
predicted by linear theory, $\delta_{\rm L} (a=1) = \delta_0$.
For the redshift of collapse combining equations (\ref{th30}) and
(\ref{th31}) we obtain
\begin{equation}    \label{th31a}
    \delta_0 = \delta_{\rm c} (a_{\rm coll})
    \frac{D(a=1)}{D(a_{\rm coll})}.
\end{equation}
Using the previously obtained results for $\delta_{\rm c}$ and appropriate
formulae for the linear growth of fluctuations (\ref{th8}), (\ref{th9}) or
(\ref{th6}), we can calculate the present linear density contrast of
fluctuation that collapsed at $z_{\rm coll}$. This relation can only be
inverted analytically in the case of $\Omega_0=1$, $\lambda_0=0$ when we
get $z_{\rm coll} = \delta_0/\delta_{\rm c} -1$. For other cases the
calculations have to be done numerically.

Using equation analogous to (\ref{th31a}) we can also calculate the
redshift of turn-around, $z_{\rm ta}$. Parameter $\delta_{\rm
c}(a_{\rm coll})$ has then to be replaced by the corresponding turn-around
value $\delta_{\rm ta}(a_{\rm ta})$ which we do not give here, but which
is calculated numerically from equation (\ref{th21}).
$\delta_{\rm ta}$ is close to unity for all models considered here.

Another interesting epoch in the evolution of an overdense region is the
onset of nonlinearity, which we characterize here by redshift $z_{\rm
nl}$. This is the time when the nonlinear density contrast given by
equation (\ref{th23}) reaches unity. Again equation equivalent to 
(\ref{th31a}) can be used with
$\delta_{\rm c}$ replaced by $\delta_{\rm nl}(a_{\rm nl})$, which turns
out to be of the order of $0.5$. We obtain $\delta_{\rm nl}$ from equation
(\ref{th21}) replacing $a_{\rm ta}$ and $s_{\rm ta}$ by $a_{\rm nl}$ and
$s_{\rm nl}$ where $a_{\rm nl}$ and $s_{\rm nl}$ obey equation
(\ref{th23}) with $\delta = 1$. 

\begin{figure}
\begin{center}
    \leavevmode
    \epsfxsize=8cm
    \epsfbox[50 30 340 790]{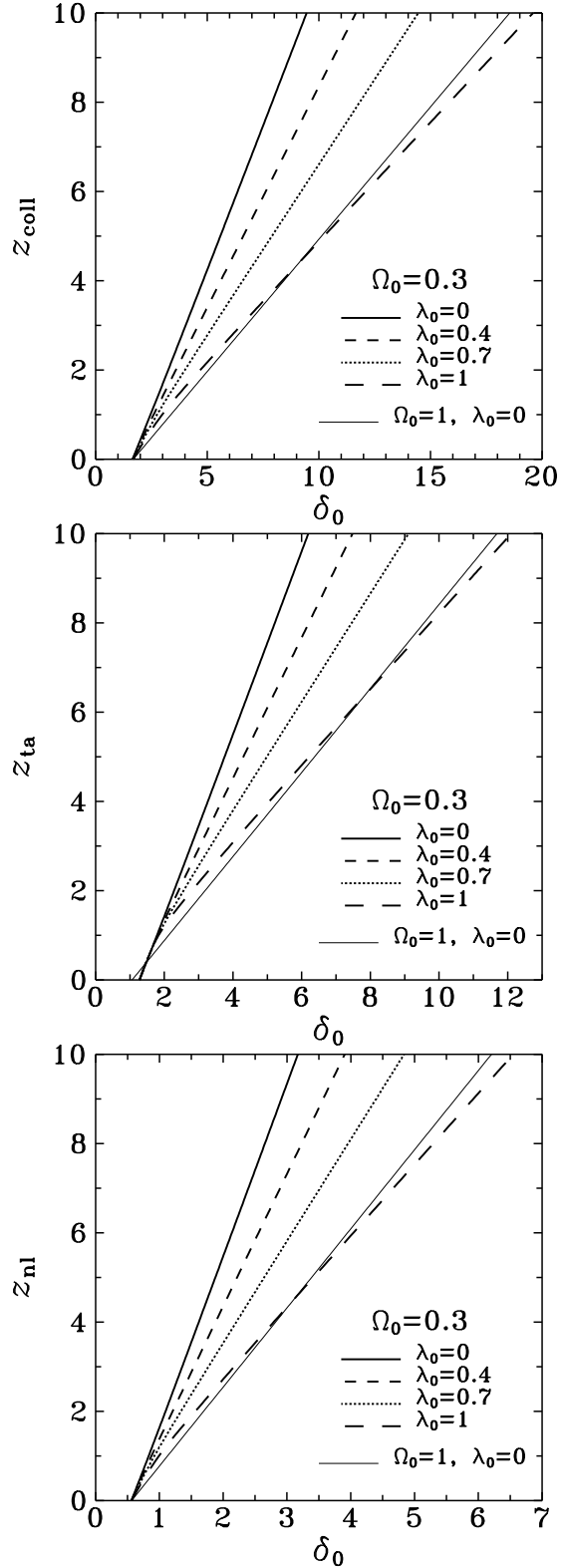}
\end{center}
    \caption{Redshift of collapse (upper panel), turn-around
    (middle panel) and transition to nonlinearity (lower panel) of density
    fluctuation with
    present linear density contrast $\delta_0$ for four models with $\Omega_0=0.3$
    and $\lambda_0=0, 0.4, 0.7$ and $1$. The thin straight line in each panel marks
    the fiducial case of $\Omega_0 = 1$, $\lambda_0 = 0$.}
\label{z}
\end{figure}

Figure~\ref{z} presents the characteristic redshifts $z_{\rm coll}$,
$z_{\rm ta}$ and $z_{\rm nl}$ as functions of $\delta_0$ for different
models with $\Omega_0=0.3$ and cosmological constant $\lambda_0=0, 0.4,
0.7$ and $1$. The thin solid line
in each panel gives for reference the exactly linear
relation in the fiducial case of $\Omega_0=1$, $\lambda_0=0$.

Results shown in Figure~\ref{z} suggest that the dependence of characteristic redshifts on
$\delta_0$ can be well fitted with simple linear formulae
\begin{equation}    \label{th35}
    z = \alpha \ \delta_0 - \beta
\end{equation}
where different constants $\alpha$ and $\beta$ correspond to each of
the three characteristic redshifts. In the $\Omega_0 =1$, $\lambda_0=0$
model this relation is exact, we have $\beta=1$ and $\alpha_{\rm
coll}=1/\delta_{\rm c}$, $\alpha_{\rm ta}=1/\delta_{\rm ta}$, $\alpha_{\rm
nl}=1/\delta_{\rm nl}$. For the most popular flat Universe with
$\Omega_0=0.3$ and $\lambda_0 =0.7$ we find the best fitting parameters $\alpha$
and $\beta$ shown in Table~\ref{fitpar}. The fits are intended to be useful at high
redshifts where the relation between $z$ and $\delta_0$ is almost exactly
linear. The accuracy of the fits in terms of $z$ obtained for a given
$\delta_0$ for $z>1$ is shown in the last column of the Table. It should
be emphasized that for high redshifts the fitted values of $\alpha$ and
$\beta$ work much better than $\delta_{\rm c}$, $\delta_{\rm ta}$,
$\delta_{\rm nl}$ and $\beta=1$.

\begin{table}
\caption{Values of the best-fitting parameters $\alpha$ and $\beta$ of
equation (\ref{th35}) for $\Omega_0=0.3$, $\lambda_0=0.7$.}
\label{fitpar}
\begin{center}
\begin{tabular}{cccc}
$z$ & $\alpha$ & $\beta$ & accuracy \\
\hline
$z_{\rm coll}$ & $\; \, 0.774$ & 1.15 & 3 \%  \\
$z_{\rm ta}$   & 1.25  & 1.30 & 5 \%  \\
$z_{\rm nl}$   & 2.32  & 1.15 & 3 \%  \\
\hline
\end{tabular}
\end{center}
\end{table}

\section{Summary}

The top hat model has been extended here to the cases of arbitrary 
positive cosmological constant including non-flat cosmologies. In
particular, we have calculated the critical (over)density for collapse,
the virial density and the characteristic redshifts of the collapse
process. These include the redshifts of the transition to nonlinearity,
turn-around and the collapse epoch. The characteristic redshifts cannot be
represented by closed analytical expressions and therefore simple
fitting formulae have been provided.

The top hat model constitutes the basic tool used in analytical and
semi-analytical models of large scale structure and galaxy formation. The
prime example here is the calculation of the mass function of collapsed
objects by the PS formalism and its recent extensions  (Lacey \& Cole
1993; Somerville \& Kolatt 1999). The PS mass function has a Gaussian term
of the critical (over)density for collapse and therefore an exact evaluation
of this density should be used. Another important application of the
top hat model is the semi-analytical modelling of galaxy formation
(Kauffmann et al. 1999; Somerville and Primack 1999). The virial
parameters of collapsed objects and the properties of baryons within
such objects are often calculated in the framework of the top hat model,
and this depends crucially on the background cosmology.

\section*{Acknowledgements}

This work was supported by the
Polish KBN grant 2P03D02319 and by the Israel Science Foundation grant
103/98.

\end{document}